# LIVELOCK FREE ROUTING SCHEMES

*Vance Faber*

**Abstract**. *We give a livelock free routing algorithm for any allowed network. Unlike some other solutions to this problem:*

*1) packets entering the network have an absolute upper bound on the time to reach their destination;*
*2) under light loads, packets are delivered to their destinations in nearly optimal time;*
*3) packets with desired paths far away from congested areas will have routing times far shorter than packets wanting to access congested areas;*
*4) if the network becomes congested and later clears, the network operates just as it would have when it was initially under a light load.*

*The main ideas of this note appear in a different form in my 1994 patent 5,369,745. This note adds to those results and makes them more mathematical.*

**Introduction**. The type of network we allow in this paper is one that can be modeled by a directed multigraph consisting of vertices representing routers and edges representing unidirectional communication links. The main restriction on the routers is that every router must have the same input capacity as it has output capacity represented by an equal number of input and output arcs. It is also assumed that the packets being routed can all be transferred from one router to a neighbor in a fixed unit of time. At each time step, the routers take packets from the input edges and route them to the output edges using a routing algorithm dependent upon the destination addresses of the packets and any knowledge of the state of the network. We assume that the number of states of the network is bounded. Packets are not allowed to exit the network except at their destinations. (This disallows a routing algorithm that throws away packets at any time.) The routers can accept new input packets only if there is a free output edge. (Note that this type of network is more general than it seems. It allows router buffers represented by loops in the graph. The number of loops is proportional to the capacity of the buffer. It can also represent networks with communication wires of varying capacity by representing these as multiple edges between routers. If it is necessary to represent the effect of distance between routers, one can interpolate intermediate "virtual" routers between two real routers to produce an allowed network.)

We give a livelock free routing algorithm for any allowed network.

Some parts of this paper appear in a totally different form in the patent [1].

**Conflict resolution in routing schemes**. Central to all routing schemes is *conflict resolution*. A conflict occurs when one or more incoming packets requests the same output destination. The router has an especially difficult decision to make if the only choice it has available is to route a packet in an undesirable direction. Some methods of conflict resolution use randomness to choose destinations. The randomized routing schemes often lead to situations where the probability of a packet not reaching its destination is extremely small but still nonzero. We do not allow that in our livelock free routing schemes; we look only at schemes where packet delivery is certain. Routing schemes often use *priority* information in order to decide the order in which choices will be made in case there is conflict. Some typical methods of conflict resolution using priority are collision counting, priority based on destination and history tables. Packets gain priority based on the state of the network and their past history in the network. Although priority is useful for resolving conflict, it is not a panacea. Once the router is faced with two packets of the highest possible priority requesting to be routed in the same direction, it has to make a somewhat arbitrary decision as to which packet request to satisfy and the other packet will not be able to increase its priority. There is the possibility that a packet will never reach its destination. There have been several methods proposed for solving this problem. Completely randomized routing [2] is known to guarantee almost certain delivery of messages but as mentioned, there is still a small probability that messages will not reach their destination in a reasonable amount of time. Besides, under light loads, randomized routing does not give optimal times. Another class of methods forces all packets of the highest priority to follow a special path that will guarantee that they reach their destination. For example, they can follow a given Eulerian path [3] through the network and there will never be collisions among them. However, in practice, once the network becomes congested all packets will be following the Eulerian path. The maximum time a packet may take before reaching its destination is heavily dependent on the construction of the Eulerian path. If the network allows it, the Eulerian path might follow repeated Hamiltonian cycles which would mean that the packets would at most have to visit all the vertices of the network once before escaping. In other cases, the time might be much longer. Finally, it has been proposed [for example, 4] that when the system reaches capacity, packets are to be thrown out. We have expressly forbidden that solution by insisting that packets can only leave the network at their destinations.

**Livelock**. A network is said to be in *livelock* when it has reached a state where

a) no packets will ever be able to exit at their destinations and
b) no new packets will ever be able to enter the network.

A routing scheme is livelock free if there is a constant *S* such that any packet entering the network is guaranteed to reach its destination in time *S*.

**Simple priority schemes**. Priority schemes maintain a value for each packet called its priority. We assume that there is an upper limit to the size this value can have. When conflicts need to be resolved, the router gives precedence to the packet with highest priority. One simple scheme is *collision counting*. Each time a packet fails to be routed to its desired destination, it receives a boost in its priority. The problem with collision counting is the danger that the value will reach its maximum and the priority will be no help in resolving conflicts. We shall give an example of that below. A more complicated version of this is to store the history of the packet, not just the number of collisions.

Again, the amount of data that is stored may reach its limit and priority can be useless in resolving conflicts. Another scheme might involve calculating priority based on the distance to the destination. The information needed for this priority is bounded. We call this *distance priority*. We show below that distance priority can lead to livelock. We propose a variant of this type of routing we *call inverse distance priority*. In this case, a packet gets priority inversely proportional to the distance to the destination. We shall show below that although inverse distance routing can lead to livelock, it has properties that are useful in creating a livelock free routing scheme.

**An example of a network in livelock**. The network in Figure 1 is in livelock when the routing scheme uses distance to the destination to assign priority. In this figure, the squares on the edges represent packets and the letters in the squares are their desired destinations. The stars mark which packets have the highest priority based on their distance to their destinations. Since each vertex has only two inputs and two outputs, once the top priority packet is assigned an output, the other packet will be assigned the other output. In this case, after one time step, although all the packets move, the figure will remain identical. If we imagine that all these packets have collision counters and those counters are at their maximum value, the fact that packets are or are not involved in collisions makes no difference in this example. Furthermore, randomizing the destinations of losing packets has no effect in this case because there is only one losing packet and there is only one output choice.

**Flushable routing schemes**. A routing scheme is *flushable* in time $T$ if given any configuration of packets in the network, when we prevent any more packets from entering the network, all packets in the network exit within time $T$. Clearly it is necessary for a livelock free routing algorithm with time bound $S$ to be flushable in time $T \leq S$.

Theorem 1. *The Eulerian routing scheme is flushable.*

Proof. This is obvious and we include it here just for completeness. The Eulerian path can be viewed as a very large cycle that contains all the edges of the network. In this case, the routers never have to resolve conflicts and there is no need for priority.

Theorem 2. *The inverse distance priority scheme is flushable*.

Proof. Suppose we are given a configuration of packets in the network and we prevent any more packets from entering. At any time later, we look at the set of packets that have the lowest distance remaining among all the packets in the network. Some one of these will have their exit port request honored and thus move to a distance state one less. In at most $k$ steps, where $k$ is the maximum distance between any two nodes in the network, some packet will leave the network. Thus, this scheme is flushable in time $T$ which is certainly no greater than the total number of packets contained in the network times the maximum distance $k$.

**Scheme promotion**. A routing scheme $\sigma$ can be *promoted* by a routing scheme $\tau$ by any priority scheme. The way to accomplish this is to have the router employ the scheme $\sigma$ until such time that the priority of a packet reaches its maximum state. After that, the scheme $\tau$ is used. When the router has to resolve a conflict, it will check to see who has the highest priority. If there are any packets for which the scheme $\tau$ is in force, the packets that are still employing $\sigma$ will have no effect on the routing. Effectively, only $\tau$ will matter.

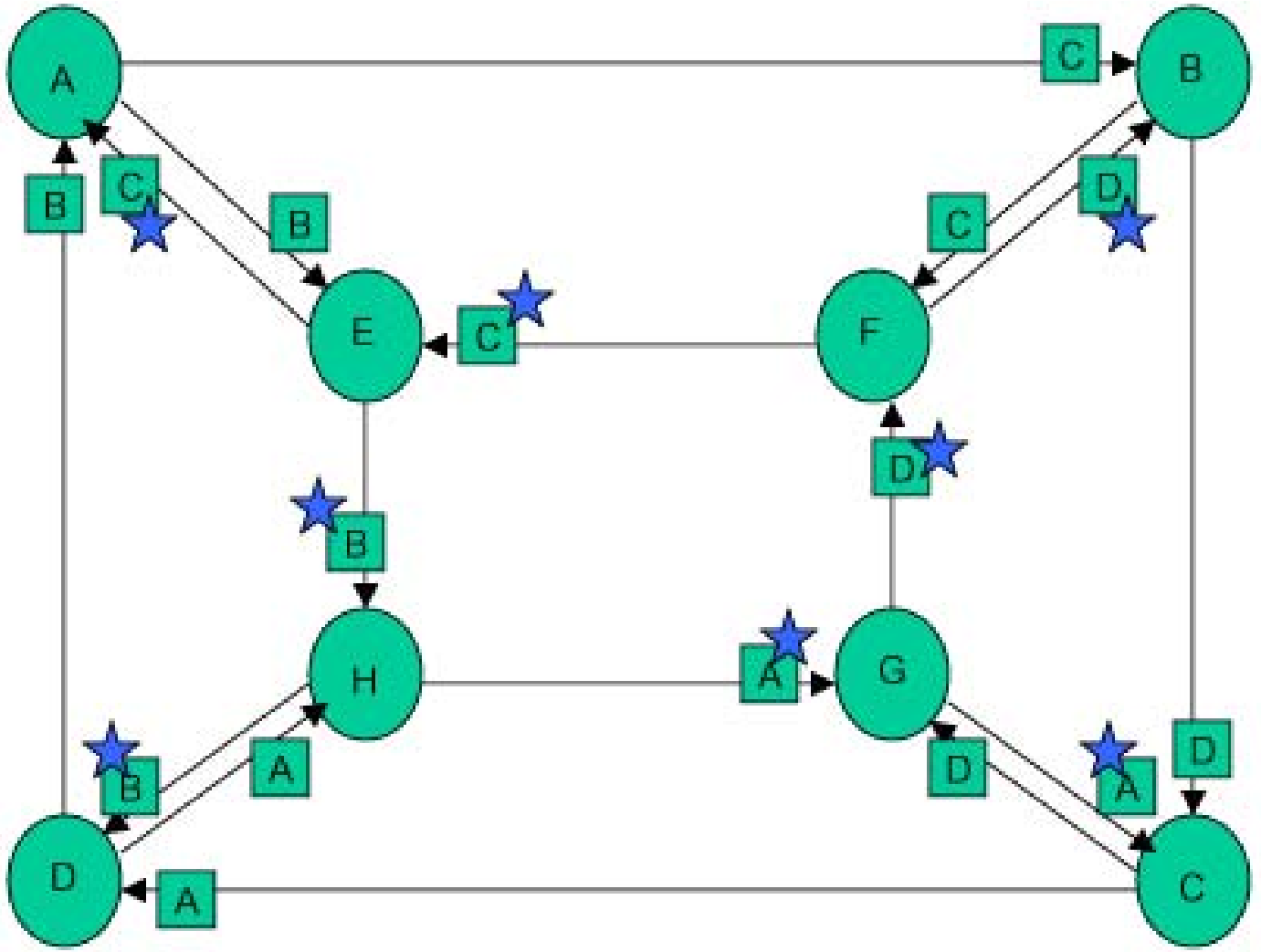

Figure 1

Theorem 3. *Every routing scheme can be promoted to a flushable routing scheme by a flushable routing scheme*.

Proof. Clearly once no packets are entering the network, either all the packets will exit or there will be conflicts. Eventually, the conflicts will cause packets to flip into the flushable state and they will then be assured of exiting the network.

Remark. For example, we could use an underlying randomized routing scheme with collision counting. After a small number of losing collisions, the packet could flip to inverse distance priority and be assured of exiting the network when no packets are entering the network.

**Paper, scissors, rock schemes**. We now describe a priority scheme that transforms *any* scheme flushable in time $T$ into one that is livelock free. The scheme involves three priority states denoted by $P$(*aper*), $S$(*cissors*), and $R$(*ock*). These states are preset as the packet enters the network by the following schedule: $R$ is assigned to all packets for $T$ clocks, then $S$ for $T$ clocks, then $P$ for $T$ clocks, then $R$ for $T$ clocks and so on. Priority is assigned cyclically: $R$ is higher than $S$, $S$ is higher than $P$, and $P$ is higher than $R$. (There may be more priorities, $D$, $E$, $F$ and so on assigned in some more complicated manner - it makes no difference in the explanation (which follows). These priorities take precedence over any other priorities and once assigned are never changed and remain constant as long as the packet remains in the network.

Theorem 3. *Paper, scissors rock schemes transform networks flushable in time T into livelock free networks.*

Proof. First note that at any given time the network can never contain packets with more than two of the priorities $P$, $S$, and $R$. This is certainly true initially when all the packets are entering with priority $R$. By the time the $P$ packets enter, the $R$ packets will have exited since there are $T$ time steps in which the $R$ packets have highest priority in the network and no more $R$ packets are entering and thus the network flushability removes the $R$ packets. Similarly, the $S$ packets will exit during the time $P$ packets are entering and before new $R$ packets enter. This continues forever. Note that newly entering packets may be bumped out of the way by old packets in the network, but within $T$ steps they will reach the highest $P$, $S$, $R$ priority and act as if the flushable scheme is the only one imposed upon them. Thus highest priority packets route normally – a distinct advantage. Lower priority packets are still free to take the shortest possible path at any time that network traffic is light.

**Variation with only two priority states**. We can also make this scheme work with only two states. We label the two states A and B. As before, we assign A for *T* clocks, B for *T* clocks, A for *T* clocks, and so on. Now, however, the routers give B top priority during the *T* clocks in which A's are being assigned and they give A top priority during the time in which B's are being assigned. Clearly, this converts any flushable scheme into one which is livelock free.

**Summary**. We have shown how with the addition of as few as two priority states, any flushable routing scheme can be transformed into one which is livelock free. This scheme satisfies these desirable properties:

1) packets entering the network have an absolute upper bound on the time to reach their destination;
2) under light loads, packets are delivered to their destinations in nearly optimal time;
3) packets with desired paths far away from congested areas will have routing times far shorter than packets wanting to access congested areas;
4) if the network becomes congested and later clears, the network operates just as it would have when it was initially under a light load.

**References**.